\documentstyle{article}
\title{
Extremely High Energy and \\ Violation of Lorentz Invariance\thanks{To be published in {\it Proc. of 
the International Workshop: Space Factory on JEM/ISS 
held on June, 1999}}
}

\author{Humitaka SATO\thanks{sato@tap.scphys.kyoto-u.ac.jp} \\ Department of Physics, Kyoto University, Kyoto606-8502, Japan}

\begin{document}

\maketitle

\begin{abstract}

Extremely high-energy(EHE) cosmic rays might provide a chance to check a violation of the Lorentz symmetry of spacetime. Some theoretical consideration is described about why the Lorentz symmetry might break-down in EHE phenomena in this universe.  Some models which introduce the violation of the Lorentz symmetry will be discussed.
\end{abstract}

\section{Violation of Symmetry in Actual Universe}

Great achievement of the 20-century Physics was discoveries of various symmetry hidden deep in the diversity of peripheral phenomena: we can count many  symmetries such as rotational and boost  symmetry of 3-space, past-future symmetry in mechanics, duality symmetry between electro- and magneto-fields, Lorentz symmetry of spacetime, discrete symmetry in atomic structure of solid, particle-antiparticle symmetry, isospin symmetry of nuclear force, chiral symmetry, "eight-fold symmetry", super-symmetry, colour symmetry and so on. Particularly, in the middle of 1970's, the pursuit to find a theory on fundamental interactions of elementary particle  focused into the unified-gauge-theory based on internal or local symmetry hidden in electro-weak and strong interactions among quarks and leptons. 

This unification of the fundamental interaction was accomplished, however, by an additional idea called "spontaneous symmetry breakdown(SSB)", which is schematically written as
$${\rm [observed~~ law]=[symmetric~~ law]x[SSB]}.$$

This SSB has introduced a new ingredient about the concept of  physics law, that is, the physics law itself is symmetric but our actual universe is not in a state of exact symmetry. This may be re-phrased also, physics law is universal but our universe is not universal entity, or, physics law itself is not affected by the actual universe where we live in. In fact, some symmetries are not exact but show a tiny breakdown, like in case of CP-asymmetry. The actual composition of cosmic matter is not particle-antiparticle symmetric in spite of CPT-symmetry in physics law itself. Following these considerations, we are tempted to think that any symmetry might be not exact in this actual universe, which might have come into existence through various spontaneous selection of non-universal parameters.  

Lorentz invariance due to the symmetry claims that there is not a preferential inertia frame; that is the central dogma of relativity  principle. However, in our universe filled with the cosmic microwave background(CMB), we can easily identified the preferential frame in which CMB is observed isotropic. CMB may be created associated with some symmetry breakdowns of the vacuum state of quantum  constituting matter of our universe. Therefore, it will be speculated that some feature of interaction might have inherited a specific selection of this SSB.  Furthermore, a recent trial to construct the unified theory of spacetime and matter based on String theory has suggested qualitatively  same origin of SSB, although the  energy scale could be quantitatively quite different.  Thus we can suppose the exact Lorentz symmetry might have been  violated in a manner of SSB in "our universe".

Lorentz symmetry, however, has been built in all fundamental concepts of  modern physics, such as  Dirac field, spin, renormalization group of quantum field theory, and so on. Therefore, violation of this symmetry should not be introduced so easily. 

One of the outcome of the relativity principle is an equality of interactions in the laboratory frame and in the center-of-mass frame. This equality has been directly proved by the accelerator experiments, up to some Lorentz factor. In the above consideration of our specific universe, the laboratory frame is identified approximately with the CMB-isotropic frame and this equality can be upgraded by the observation of EHE cosmic rays including EHE cosmic neutrinos. In this respect, the Greisen-Zatepin-Kuzumin(GZK) cut-off[1][2] is in an unique status.
 
\section{High-energy tests of Lorentz symmetry}

The argument of the GZK cut-off is based on the Relativity Principle, that claims an equality of all the inertia frames. Therefore, a collision between a CMB photon of $10^{-3}$eV and a EHE proton of $10^{20}$eV in the laboratory frame or U-frame(the universe frame) is identical with a collision between $\gamma$-ray of 300MeV and a proton at rest, which is approximately the center-of-mass system. The U-frame would be identified also with the so-called comoving frame in the FRW universe model. 

Denoting the inertia frames by the Lorentz factor, $\gamma$, with respect to the U-frame, the relativity principle has been verified experimentally up to about $\gamma \sim 10^6$ by the accelerator experiment. And, if the GZK cut-off is checked in future, the  verification will be increased up to $\gamma \sim 10^{12}$. Thus the GZK cut-off is closely related with the direct verification of the Relativity Principle as pointed out by our paper in 1972[3]. Recent experimental indication of neutrino mass may also upgrade the maximum $\gamma$ of the direct verification of the Relativity Principle.

This can be formulated as follows: suppose an existence of an universal four vector $ \vec{N}$, which takes $(1,0,0,0)$ everywhere in the U-frame[4]. The energy relative to the U-frame is given as $\vec{N} \cdot \vec{P}$, $\vec{P}$ being four momentum. Then  a collision cross-section between two particles with $\vec{P_{(1)}}$ and $\vec{P_{(2)}}$ could be  expressed generally as[5]
   $$ \sigma(Q,P_{(1)},P_{(2)})$$
,where $Q$ is a relative four momentum, $Q^2=(\vec{P_{(1)}}+\vec{P_{(2)}}) \cdot (\vec{P_{(1)}}+\vec{P_{(2)}})$ and $P_{(a)}=\vec{P_{(a)}}\cdot \vec{N}$. Relativity Principle requires that the cross-section $ \sigma $ does not depend on   $ P_{(a)} $ and must be
         $$\sigma (Q). $$

If $\sigma$ would depend on $ P_{(a)} $, new exotic effect could be introduced: suppose that a density of state is  suppressed by some reason for such momentum $\vec{P}$ that 
$$ P_{(a)}  > \sqrt{p_c^2 + m^2} ~~~ {\rm for}~a=1,2$$,
where $p_c$ is a cut-off momentum. Under such assumption, the suppressed region of the momentum space is relocated in the non U-frame. Then, in the extreme high $\gamma$-frame, the origin in that frame is shifted into the suppressed region for $ \gamma > p_c/m $ and even a low energy interaction in the center-of-mass system turns into the suppressed region because the final state density is suppressed. Thus the GZK cut-off would disappear if $p_c/m < 10^{12}$[3].

\section{Violation of Lorentz invariance}
How does $P_{(a)}$ come in the $\sigma$? That could happen by introducing the Lorentz-violating term  in the Lagrangian. We give some illustrative examples in the followings.

(a) Shift of light velocity[6][7]

There are two origin of special relativity: One is the  mechanical equivalence among inertial frames for a ponderable matter, which introduces a limiting velocity $c_{\rm m}$. Another one is the electrodynamic relativity, which introduces the constant light velocity, $ c_{\rm em}$, in any inertia frames. Combining these two physical contents, the action for a charged particle is written as 

$$ I= \int dt \left[ -mc_{\rm m}^2(1-{ v^2 \over c_{\rm m}^2})^{1 \over 2} + { e \over c_{\rm m}}A_\mu v^\mu \right] + {1 \over 8 \pi} \int dx^3 dt \left[ {\bf E}^2 -({ c_{\rm em} \over c_{\rm m}})^2{\bf B}^2 \right]  $$

Einstein's special relativity implies simply $c_{\rm m}=c_{\rm em}$ and the Lorentz invariance holds for the Lagrangian. The electromagnetic part of the above Lagrangian can be rewritten as
$$  {1 \over 8\pi} \left[ {\bf E}^2 -({ c_{\rm em} \over c_{\rm m}})^2{\bf B}^2 \right] ={1 \over 8\pi} \left[ {\bf E}^2 -{\bf B}^2 \right]+ \epsilon {1 \over 8\pi}{\bf B}^2 $$
with $ \epsilon=(1-({ c_{\rm em} \over c_{\rm m}})^2)$. The last term in the right hand side is not Lorentz invariant and various experimental check has constrained the coefficient $\epsilon $ as  $ \epsilon < 10^{-21}$ .

(b) Vacuum Cherenkov radiation and Photon decay

Coleman and Glashow[8] pointed out that an exotic channel of particle processes opens in case of $c_{\rm m}\neq c_{\rm em}$ and high energy check of Lorentz violation will be possible: If $c_{\rm m}<c_{\rm em}$, the photon decay $ \gamma \rightarrow e^+ + e^-$ is    
opened above the the threshold energy $E=2m/\sqrt{  ( c_{\rm em}/c_{\rm m})^2-1}$, and , if $c_{\rm m}>c_{\rm em}$, the vacuum Cherenkov radiation of charged particle become possible for the energy above $ m/\sqrt{1-(c_{\rm em}/c_{\rm m})}$.

They conjectured furthermore that the velocity eigen state of massless particle could be different from the  limiting velocity $c_{\rm m}$. For neutrinos, if the flavor eigen state and the velocity eigen state is not identical, the neutrino oscillation would happen.

(c) Energy dependent light velocity

If we  modify the conventional relation of $ c_0^2 {\bf p}^2=E^2$ as[9]

$$ c_0^2 {\bf p}^2\approx E^2 [1+f(E)]$$

,where $E$ is the energy in the U-frame. Since the modification might be due to an effect of Quantum gravity, $f(E)$ term would dominate in the Planck energy $E_{Pl}$ and, in the low energy limit, it will take a form of $ f(E) \approx \zeta (E/E_{Pl})$. Then the velocity becomes

 $$ c={ \partial E \over \partial p}\approx c_0(1-\zeta {E \over E_{Pl}})$$ 
 Combining the stability argument of high energy photons in the above (b), the requirement   to explain observed TeV gamma-ray from astronomically  remote objects can derive the constraint for $|1-(c_{\rm em}/c_{\rm m})|$[10] . 

Interests on the photon mass has revived in various contexts recently[11][12].
In case of finite mass, the four vector potential itself has a physical meaning in different from the conventional gauge-potential. Then it might be paved 
the all space in our universe and might work to provide  the universal four vector.
 
(d)Lorentz-Violating Interaction terms

In the case of (a), the new exotic channel such as discussed in (b) opens abruptly above the threshold  energy,  in contrast with a gradual effect with higher energy in case of (c). Coleman and Glashow[13] have pointed out that the Lorentz violating interaction term introduces "velocity mixing" of the particle at the high energy in the U-frame; by adding the violating term, the conventional Lorentz invariant Lagrangian is modified as

$$ L \rightarrow L + \partial_i {\bf \phi} {\bf \epsilon} \partial^i {\bf \phi} $$
,where $i=1,2,3$,~~ ${\bf \phi}$ is a set of fields  and  ${\bf \epsilon}$ is a matrix which does not commute with the mass matrix. By this modification, the single-particle energy-momentum eigenstates changes from the eigenstates of  the mass matrix at low energy into the eigenstates of ${\bf \epsilon} $ at high energy.

The equivalent modification can be written also as 

$$  L \rightarrow L + \partial_0 {\bf \phi} {\bf \epsilon} \partial^0 {\bf \phi}.$$

 How such term could happen spontaneously from the Lorentz-invariant form? We speculate that such terms could happen from the Lorentz invariant term via SSB. For example, suppose such Lorentz invariant term  as below
$$  \left[ \partial_{\mu} {\bf \phi} \partial^{\mu} \Phi \right] {\bf \xi} \left[\partial_{\mu} \Phi \partial^{\mu} {\bf \phi} \right].$$

Here, $\Phi$ is a set of  fields which associate with determining the vacuum state of the constituting particles of the present universe and can not be excited locally at the present as the ordinary particles. Inflaton fields and fields determining the cosmological vacuum term or so-called Quintessence are  classified to this kind of fields. Those fields are now in the spatially uniform state. Then,  the above term reduces into the Lorentz violating term with 
$$ {\bf \epsilon}=\partial^0 \Phi {\bf \xi}\partial_0 \Phi. $$
Here $\Phi$ is supposed to be changing only with cosmological time-scale, that is, such as  $\partial_0 \Phi \sim H_0 \Phi $, $H_0$ being the Hubble constant at present. Therefore, we can easily expect the smallness of $\epsilon$, even if $\xi$ is not exceptionally small.

 {\bf References}

\vspace{1pc}

 1. Greisen, K., 1966, {\it Phys. \ Lett.} {\bf 16}, 148.      

 2. Zatsepin, G. T. and Kuz'min, V. A., 1966, {\it JETP \ Phys.\ Lett.} {\bf 4}, 78.

 3. Sato, H. and Tati, T., 1972, {\it Prog. \ Theor. \ Phys.} {\bf 47}, 1788.

 4. Blokhintsev,~~D.~I., 1966, {\it Sov. \ Phys. \ USPEKHI}~~{\bf 9}, 405.

5. Sato, H., 1998, {\it Black Holes and High Energy Astrophysics}  edited by 
H. Sato and N. Sugiyama, Universal Academy Press, 401.

6. Haugan, M. P. and Will, C. M., 1987, {\it Physics Today } {\bf May}, 69.

7. Green, G. L., Dewey, M. S., Kessler, E. G., and Fischbach, E., 1991,{\it Phys. Rev. } {\bf D44}, 2216.

8. Coleman,~S. and Glashow,~S., 1997, {\it Phys. \ Lett. } {\bf B405}, 249.

9. Amelino-Camelia, G., Ellis, J., Mavromatos, N. E., Nanopoulous, D.V., and Sarkar, S.,1998, {\it Nature } {\bf 393}, 763.

10. Kifune, T., astro-ph/9904164, Ap.J(in press).

11. Lakes, L., 1998, {\it Phys. \ Rev. \ Lett.} {\bf 80}, 1826.

12. Schaefer, B. E., 1999, {\it Phys. Rev. Lett.} {\bf 82}, 4964.

13. Coleman, S., and Glashow, S. L., 1999, {\it Phys. Rev. } {\bf D59}, 116008.

\end{document}